\documentclass{aastex62}
\usepackage{subfigure}
\graphicspath{{./}{figures/}}

\shorttitle{Modified Gravity at Astrophysical Scales}
\shortauthors{Cerme\~no et al.}

\begin{document}

\title{Modified Gravity at Astrophysical Scales}

\author{M. Cerme\~no}
 \email{marinacgavilan@usal.es}
 \affiliation{Department of Fundamental Physics, University of Salamanca, Plaza de la Merced s/n 37008 Salamanca, Spain
}  
\author{J. Carro}
\affiliation{Departamento de F\'{\i}sica Te\'orica and UPARCOS, Universidad Complutense de Madrid, 28040 Madrid, Spain}
 \author{A. L. Maroto}
 \email{maroto@ucm.es}
\affiliation{Departamento de F\'{\i}sica Te\'orica and UPARCOS, Universidad Complutense de Madrid, 28040 Madrid, Spain}
\author{M. A. P\'erez-Garc\'ia }
\email{mperezga@usal.es}

\affiliation{Department of Fundamental Physics, University of Salamanca, Plaza de la Merced s/n 37008 Salamanca, Spain
}

\begin{abstract}


Using a perturbative approach we solve stellar structure equations for low-density (solar-type) stars whose interior is described with a polytropic equation of state in scenarios involving a subset of modified gravity theories. Rather than focusing on particular theories, we consider a model-independent approach in which deviations from General Relativity are effectively described by a single parameter $\xi$. We find that for length scales below those set by stellar General Relativistic radii the modifications introduced by modified gravity can affect the computed values of masses and radii. As a consequence, the stellar luminosity is also affected. We discuss possible further implications for higher density stars  and observability of the effects before described.
\end{abstract}


\section{Introduction}

It is well established that General Relativity (GR) provides an accurate  description 
of the gravitational interaction from the sub-millimeter scales probed by torsion balance 
experiments \citep{torsion, torsion2} to kiloparsec distances as confirmed by recent observations of 
strong gravitational lensing of extragalactic objects \citep{lensing}. These tests probe the weak-field regime of the theory and so far no discrepancy with respect to the 
predictions of GR has been found. For strong fields, the theory is 
still poorly tested, but the recent discovery of gravitational waves produced in the 
merger processes of black holes \citep{GWBH} or neutron stars \citep{GWNS} has opened a new avenue to explore this regime of the theory. Despite the success of GR on small (sub-galactic) scales,
the difficulties to accommodate the observed accelerated expansion of the universe within
the theory has led to the suggestion that
the universal attractive character of gravity could break down on cosmological scales. Several Modified Gravity (MG) theories have been proposed in the last years which  introduce additional degrees of freedom, typically scalars, that mediate the gravitational interaction thus changing its behaviour on very large scales.

On small scales, the new degrees of freedom are 
usually screened \citep{k1,k2, screening} so that the agreement of standard GR with observations
is not spoiled \citep{j1,j2}. However,  
in certain modified gravities, such as beyond Horndeski models \citep{GPLV1,GPLV2},  the screening mechanism is only partially operational, and, in particular, it could break down \citep{MS1,MS2} inside 
astrophysical objects, such as stars, where a weakening of the
gravitational interaction is predicted \citep{BD}. This possible modification has also been studied in other approaches \citep{martinsl}. Such deviations from Newton's law can produce modifications in the internal stellar structure. This fact has motivated the use of
different types of stars i.e main sequence \citep{MS1,PRL1,Velten}, white dwarfs \citep{Babichev:2016jom,WD1,PRL2} or neutron stars
\citep{Babichev:2016jom,Velten, capoz} as probes for alternative gravitational models.

Despite the fact that some of the beyond Horndeski 
theories have been practically ruled out by the observation of the GW170817 event, 
still some of them are viable modifications of gravity \citep{Zuma1,Zuma2}. For a discussion see \citep{p1,p2,p3}.
Thus, in such  theories the equation for hydrostatic equilibrium in the non-relativistic regime is modified in such a way that \citep{MS1}
\begin{eqnarray}
\frac{d P}{dr}=-\rho\frac{Gm(r)}{r^2}-\frac{\rho\Upsilon G}{4}\frac{d^2m(r)}{dr^2},
\label{eqbH}
\end{eqnarray}
where $P$ is the pressure, $m(r)$ is the mass inside a radius $r$, $\rho$ is the energy density, G is the gravitational constant \citep{pdg} and $\Upsilon$ is a constant parametrizing the deviation with respect to GR. 
Existing bounds $-0.22<\Upsilon<1.6$ are set by the Chandrasekhar limit on white dwarf stars \citep{PRL2,Babichev:2016jom} and the minimum mass of main sequence stars \citep{PRL1,Babichev:2016jom,saltas}. 

On more general grounds, effective descriptions of alternative gravity theories have been 
developed in recent years  aiming at encapsulating in a few parameters all the relevant 
modifications at a given scale \citep{PF, silv}. These parametrizations have
been widely employed in the analysis of structure and lensing data of galaxy surveys \citep{surveys}. 

In this work we will explore the implications for stellar structure of one of the 
simplest and widely used effective parametrizations considered in the literature \citep{silv,bert}. For a wide class of theories of gravity with one additional scalar degree of freedom, assuming the quasi-static approximation and not higher than second order equations of motion, it can be seen that all the relevant modifications can be encoded in an effective Newton's constant parametrized in Fourier space by $\mu(k)=G_{\rm eff}(k)/G$ with 
$G$ the ordinary Newton's constant, and a gravitational slip parameter $\gamma(k)=\phi(k)/\psi(k)$. The $\mu(k)$ parameter changes the hydrostatic equilibrium equation  introducing, generically, a new length scale in the dynamics and leading to expressions which can 
deviate from Eq. \ref{eqbH}. This effective approach will allow us to analyze the potential modifications in the stellar structure, including radius, mass, luminosity and temperature  of stars, in a largely model-independent way. 

The manuscript is structured as follows. In Sec. 2, we briefly review and expose the main features of the modified gravity theories we consider in our analysis of the changes  induced in the stellar structure equations.  We also present the main scenario we explore i.e. low density stars and discuss the validity of our approach. The results are then detailed in Sec. 3, and final conclusions are drawn in Sec. 4.

\section{Stellar structure equations}

In this section we start by presenting the standard set up of the calculation for the structure of a stellar object of mass $M$ and radius $R$ in General Relativity. We consider for the sake of simplicity a non-relativistic and non-rotating object. Following \citet{silv} we consider a spherical Minkowski metric with linear scalar perturbations through the introduction of two radial potentials, $\psi(r)$ and $\phi(r)$, both fulfilling the condition of weak field approximation i.e. $\psi(r)\ll 1$, $\phi(r)\ll 1$. The interval $d\tau^2=g_{\mu \nu}dx^\mu dx^\nu$ is written as
\begin{equation}
d\tau^2=-(1+2\psi(r))dt^2+(1-2\phi(r))(dr^2+r^2d\Omega^2),
\end{equation}
where 4-coordinates $x^\mu$ are $t,r,\theta,\phi$.

From the selected metric, $g_{\mu \nu}$, and using the approximation of a perfect fluid,  the energy momentum tensor can be written as
\begin{equation}
T_{\mu \nu}=(\rho+P)U_\mu U_\nu+Pg_{\mu \nu}.
\end{equation} 
$U^\mu$ is the fluid four-velocity. We assume static solutions with $U_\mu=(U_t,0,0,0)$, and impose $U^\mu U_\mu=-1$. We can obtain the Einstein's equations $G_{\mu \nu}=8\pi G T_{\mu \nu}$ with $G_{\mu \nu}=R_{\mu \nu}-\frac{1}{2}g_{\mu \nu}R$ the Einstein's tensor given by the Ricci tensor, $R_{\mu \nu}$, and the Ricci scalar, $R$, for this metric. Explicitly, these expressions read to first order in metric perturbations 
\begin{equation}
2\phi'(r)+r\phi''(r)=4\pi G r \rho(r),
\label{1}
\end{equation}
\begin{equation}
(\psi(r)-\phi(r))'=4\pi G r P(r)
\label{2}
\end{equation}
and
\begin{equation}
(\psi(r)-\phi(r))'+r(\psi''(r)-\phi''(r))=8\pi G r P(r).
\label{3}
\end{equation}
In addition, for the spatial diagonal component of the energy momentum tensor we can approximate to first order in metric perturbations $T_{ii}=P(1-2\phi)\simeq P$, being $i=1,2,3$ a spatial index.\\

As obtained, it is clear that Eq. \ref{1} is the Poisson equation for $\phi$,
\begin{equation}
\nabla^2\phi=4\pi G \rho.
\label{Poisson}
\end{equation} 
If we take into account the mass relation 
\begin{equation}
\frac{dm}{dr}=4\pi r^2 \rho ,\label{mass}
\end{equation}
then from Eq.\ref{Poisson} we obtain
\begin{equation}
\frac{d\phi}{dr}=\frac{Gm(r)}{r^2}.
\label{4}
\end{equation}
On the other hand, using Eq. \ref{2} and Eq. \ref{4} we obtain
\begin{equation}
\frac{d\psi}{dr}=\frac{Gm(r)+4\pi G r^3 P}{r^2}.
\label{TOV1}
\end{equation}
Now, if we use the continuity equation $\nabla_\mu T^{\mu \nu}=0$, we finally obtain
\begin{equation}
\frac{dP}{dr}=-(\rho+P)\frac{d\psi}{dr}.
\label{TOV2}
\end{equation}
It is important to note that Eq.\ref{TOV1} and Eq.\ref{TOV2} along with the mass equation Eq.\ref{mass} are the structure equations for the potential, pressure and mass in the star.
Besides, using Eq.\ref{2} and Eq.\ref{3} we get $\psi-\phi=C_1+C_2r^2$, with $C_1$ and $C_2$ constants. If we impose the existence of a finite solution when $r\rightarrow \infty$, it follows $C_2=0$. On the other hand, for $r> R$ we have to recover the Schwarzschild metric, which means $\psi= \phi$. In this way, $C_1=0$ and $\psi-\phi=0$, making $P=0$. This result is due to the fact that we obtain a first order perturbation solution, being $\psi\sim \phi \sim \frac{-Gm(r)}{r}$. Then, whereas $\rho$ and $\psi$ are first order functions in perturbation theory, $P$ is a second order function in perturbations. 
We can thus rewrite Eq.\ref{TOV1} and Eq.\ref{TOV2} as
\begin{equation}
\frac{d\psi}{dr}=\frac{Gm(r)}{r^2}
\label{TOV3}
\end{equation}
and
\begin{equation}
\frac{dP}{dr}=-\rho\frac{d\psi}{dr},
\label{TOV4}
\end{equation}
retaining only the leading contributions, i.e. first and the second order respectively. From Eq.\ref{TOV3} the mass of the star can be solved as
\begin{equation}
m(r)=\frac{r^2}{G}\psi'(r).
\label{TOV5}
\end{equation}

\subsection{Input from Modified Gravity theories}

In the context of MG theories and for static configurations following \citet{silv} we introduce two functions, $\mu(k)$ and $\gamma(k)$, in the Fourier $k$-space whose effect is modifying the equations governing the solution of the potentials $\psi,\phi$. We can write
\begin{equation}
\nabla^2 \psi =4\pi G \mu \rho
\label{ref3}
\end{equation}
and
\begin{equation}
\phi=\gamma \psi.
\label{gamma}
\end{equation}
The physical meaning of the $\mu(k)$ function is that of providing an effective value of the gravitational constant, $G$. Instead, $\gamma(k)$ establishes a relationship between the two  potentials, $\psi,\phi$. When $\mu=\gamma=1$ we recover GR equations. The most general expression for $\mu(k)$ in theories with one extra scalar degrees of freedom and modified Einstein's equations involving up to second order derivatives can be cast into the rational form \citep{silv}
\begin{equation}
\mu(k)=\frac{1+p_3k^2}{p_4+p_5k^2}.
\label{mu}
\end{equation}

The previous step stems from the fact that it is indeed equivalent to rewrite Eq.\ref{ref3} as
\begin{eqnarray}
p_4\nabla^2\psi-p_5\nabla^4\psi=4\pi G(\rho-p_3\nabla^2)\rho,
\label{mar}
\end{eqnarray}
which is a local expression. Having this in mind we can now obtain the expression of $\mu$ in position space as
\begin{equation}
\mu(r)=\frac{1-p_3\nabla^2}{p_4-p_5\nabla^2},
\label{mu2}
\end{equation}
with $p_3,p_4,p_5$ constant parameters. Thus, for example, by defining $\beta_1=p_3/p_5$ it has been shown \citep{bert,gianan} that several models such as
$f(R)$ or certain Chameleon theories correspond to 
$\beta_1=4/3$. The parameter values corresponding to other models like Yukawa-type theories belong to interval $0.75<\beta_1<1.25$. Furthermore, as we will show below, $\gamma(k)$ plays no role in our study, so that we do not provide any particular parametrization for it. In any case,  we should keep in mind that more general parametrizations exist \citep{hoj} which could also include additional vector degrees of freedom \citep{Aparicio}.

As can be readily seen, under the conventions used, $p_3$ and $p_5$ have units of squared length, while $p_4$ is dimensionless. Using Eq.\ref{mu2} we can rewrite Eq.\ref{ref3} as
\begin{equation}
\psi''+\frac{2}{r} \psi'=4\pi G \left(\frac{1-p_3\nabla^2}{p_4-p_5\nabla^2} \right)\rho,
\label{poi}
\end{equation}
which generalizes the Poisson equation in our post-Newtonian scenario \citep{pani}. 

As we want to consider small perturbations from GR, in what follows we demand $p_4$ does not depart largely from unity and $p_3\nabla^2\ll 1$ and $p_5\nabla^2\ll 1$. In this scenario, for spherically symmetric configurations, Eq. \ref{mass} now takes the form
\begin{eqnarray}
\frac{dm}{dr}=\frac{r^2}{G}\mu^{-1}\left[\frac{1}{r^2}(r^2\psi')'\right]\label{dmdr},
\end{eqnarray}
where the differential operator $\mu^{-1}$  is given to first order in the perturbative expansion by
\begin{eqnarray}
\mu^{-1}=p_4(1-\xi \nabla^2),
\label{mueq}
\end{eqnarray}
where 
\begin{eqnarray}
\xi=-p_3+\frac{p_5}{p_4}.
\end{eqnarray}

From integration of Eq.\ref{dmdr} we get to first order
\begin{equation}
m(r)=\frac{p_4}{G}r^2\psi'(r)-\frac{p_4\xi}{G}r^2\left(\frac{1}{r^2}(r^2\psi')'\right)'.
\end{equation}
In addition, we can also write
\begin{eqnarray}
\frac{d\psi}{dr}=\frac{Gm(r)}{p_4 r^2}+\frac{\xi G}{p_4}\left(\frac{m''(r)}{r^2}-\frac{2m'(r)}{r^3}\right),
\end{eqnarray}
which is the modified version of Eq.\ref{TOV3}. 
Finally,
\begin{eqnarray}
\frac{d P}{dr}=-\rho\frac{Gm(r)}{p_4 r^2}-\frac{\xi\rho G}{p_4}\left(\frac{m''(r)}{r^2}-\frac{2m'(r)}{r^3}\right).
\end{eqnarray}

At this point it is important to notice that the subset of MG models  we are  considering only hold in the linear regime, which we assume to be the one 
 valid inside solar-type stars (and other high-density stars such as white dwarfs). In this sense, we must emphasize that some  proposed modifications of gravity on cosmological scales which rely on  screening mechanisms  \citep{MS1,MS2} through  non-linear effects on astrophysical scales are not covered by this approach.

Notice that because of the form of the effective operator $\mu$, the structure of the pressure equation is different from that shown in Eq.\ref{eqbH}. A few comments are in order. For simplicity we have limited ourselves to the case in which the parameters $p_3$, $p_4$ and $p_5$ are just object-independent constants. 
In the most general case, for the spherically symmetric configurations considered in the work and unlike the cosmological case discussed in \cite{silv}, these parameters could be general  functions of the stellar radius.  Thus for example, the $p_4$ parameter related to the effective Newton's constant could interpolate between different values in the interior and exterior of the star. However in our case with constant $p_4$, matching with the laboratory measured value of Newton constant  $G$, imposes $p_4=1$. Thus the        combination  $G/{p_4}$ appearing in the obtained structure equations inside the stellar body will effectively take the value $G$. The possibility of changing the effective Newton's constant in astrophysical objects as compared to the laboratory value has been  considered phenomenologically in previous works in non weak-field limits, see for  instance \cite{Velten} and references therein, where an $\alpha$ parameter is used analogously to our $p_4$ or in cosmological contexts, see \cite{planck} for explicit measurements of $\mu_0$.

\subsection{Perturbative solution for a polytropic star}

We have seen that the potential $\psi$, the density $\rho$ and the pressure $P$ are related through Eq.\ref{TOV4}. In this way, given an equation of state (EoS) determining a relation between $P$ and $\rho$ we could, in principle, obtain an expression for $\psi(\rho)$.\\
In this work we use a polytropic EoS which can be considered a reasonable description for main-sequence (solar-type) stars \citep{hansen}. In addition, we will also consider the higher density case of white dwarfs. No dark matter presence is assumed \citep{cermeno1,perez-silk-daigne} in this context. As mentioned, a generic polytropic EoS takes the simple form
\begin{equation}
P=K \rho^{1+\frac{1}{n}},
\label{poly}
\end{equation}
where $n$ is the so-called polytropic index, which is related to the internal constituents of the star, and $K$ is a constant with appropriate units. In this work we use CGS units.\\
Therefore, using Eq.\ref{TOV4} and Eq.\ref{poly}, for $n>0$ we have
\begin{equation}
\rho=\left(\frac{-\psi}{(n+1)K} \right)^n, 
\label{poly2}
\end{equation}
whereas for $n=0$ the equation takes simply the form $\rho=\rho_c$,  being $\rho_c$, the central density of the star. The equation for the $\psi$ potential can be written under the form
\begin{equation}
\psi''+\frac{2}{r}\psi'=4\pi G \mu \left( \frac{-\psi}{(n+1)K} \right)^n.
\label{dpsi}
\end{equation}

In order to solve the differential equation in Eq.\ref{dpsi} using a perturbative approach, one can propose a solution under the form $\psi=\psi_0+\psi_1$, with $|\psi_1|\ll|\psi_0|$. Thus, a set of  two differential equations is obtained, one for each component, $\psi_0$, $\psi_1$.
\begin{equation}
\psi''_0+\frac{2}{r}\psi'_0={4\pi G}\left(\frac{-\psi_0}{(n+1)K} \right)^n, 
\label{ED1}
\end{equation}
and
%
%
\begin{equation}
\psi''_1+\frac{2}{r}\psi'_1=\frac{4\pi G}{(n+1)^n K^n} \left[ \xi\nabla^2(-\psi_0)^n-n(-\psi_0^{n-1})\psi_1  \right].
\label{ED2}
\end{equation}

Let us mention that this same result would follow from the alternative calculation using a perturbative expansion of  $\psi$ and $\rho$ in Eq.\ref{mar}.  Notice, that the perturbative approach allows to solve separately the gravitational potential equation from the conservation equation shown in Eq. \ref{TOV4} which remains unchanged with respect to the ordinary GR case.

In order to calculate the radius $R$ and mass $M$ we must first obtain the solution $\psi(r)$, which will provide us the density $\rho$, through Eq.\ref{poly2}. By imposing $\psi(R)=0$, the radius $R$ will be calculated from the usual consideration that both pressure and density vanish for $r=R$. 

Accordingly, the total mass of the star, $M(R)$ is obtained using Eq. \ref{poly2} and integrating Eq. \ref{mass}, so finally  
\begin{equation}
M(R)=\int_0^R 4\pi r^2 \rho dr.
\end{equation}
It is important to note that if we impose $\xi=0$ we will recover the GR case. In order to be more definite we now particularize to the perturbative solution for a polytropic $n=3$ star focusing on the case of a low mass solar-type stars $M\lesssim M_\odot$. Most of the internal description is reasonably well described by a polytrope with a $n=3$ index. Refinements to this description would involve, at least, the use of several polytropes or, more appropriately, an improved version \citep{hansen} of the Standard Solar Model (SSM) \citep{ssm}. Note however, that the goal of our calculation is not a detailed modelling of the stellar object but rather showing the effects of selected MG models, thus we will restrict to a single polytropic EoS. Therefore, in this case it can be written as
\begin{equation}
P=K \rho^{\frac{4}{3}}.
\end{equation}
In addition, Eqs.\ref{ED1} and \ref{ED2} for this specific polytrope take the form
\begin{equation}
\psi''_0+\frac{2}{r}\psi'_0=\frac{-\pi G }{16} \left( \frac{\psi_0}{K} \right)^3 ,
\label{ED1P}
\end{equation}
and
\begin{equation}
\psi''_1+\frac{2}{r}\psi'_1=\frac{-\pi G}{16 K^3}\left[ \xi\nabla^2 (\psi_0)^3 +3 (\psi_0)^2\psi_1 \right].
\label{ED2P}
\end{equation}
The density profile and the mass can be obtained as
\begin{equation}
\rho(r)=\left( \frac{-\psi}{4K}\right)^3, 
\end{equation}
from which the radius $R$ is derived through the relation it must fulfill, $\rho(R)=0$, which is needed to calculate the stellar mass
\begin{equation}
M=\frac{-\pi}{16 K^3}\int_0^R \psi^3(r) r^2dr.
\end{equation}
%
\begin{figure}[t]
\begin{center}
\includegraphics [angle=0,scale=0.85] {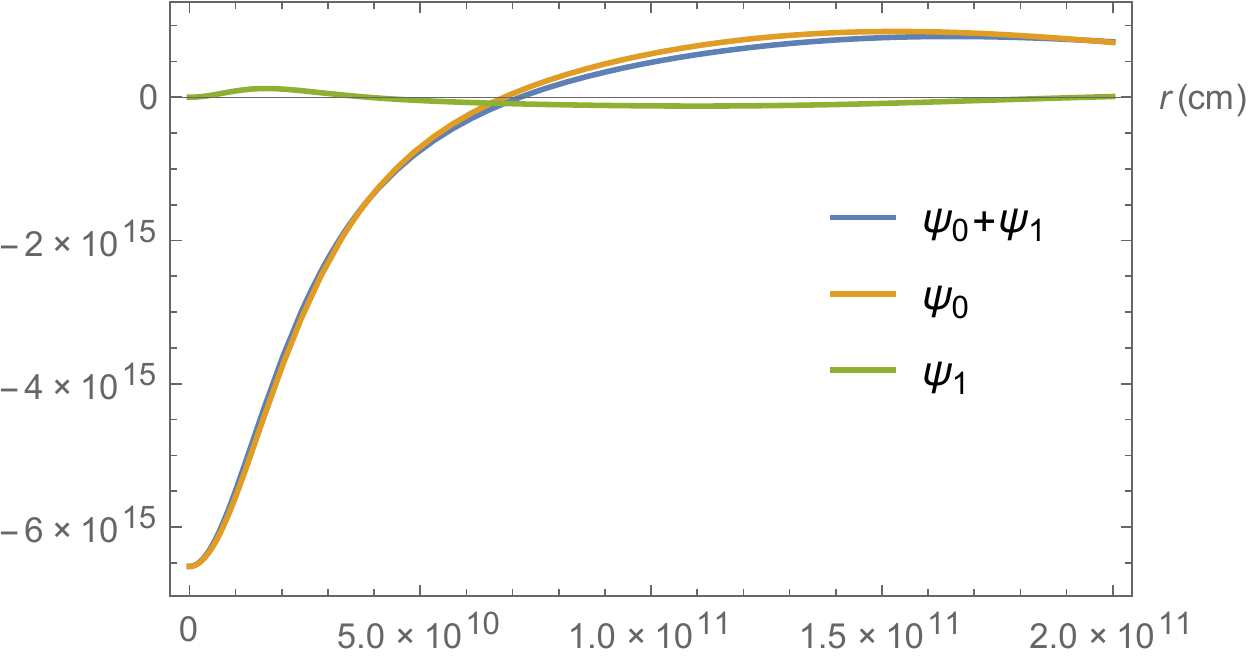}
\caption{Potential $\psi=\psi_0+\psi_1$, constructed from the unperturbed potential $\psi_0$ and the perturbation $\psi_1$ as a function of $r$ for $\xi=-2.67 \times 10^{18}$ cm$^2$. We take $\rho_c=\rho_0=80\,\rm g/cm^3$. }
\label{fig1}
\end{center}
\end{figure}
For reference, solutions for the radii and masses of solar-type configurations in the GR case we consider in this work have been obtained using $K=3.8 \; \rm {cm^3}{g^{-\frac{1}{3}}s^{-2}}$ so that we can recover values close to those of the sun, $M_0=1.95674\times 10^{33}$ g and $R_0=6.8141\times 10^{10}$ cm. A central density value $\rho_c=\rho_0=80\,\rm g/cm^3$ has been used.

In addition, stellar luminosity, $L$, can be derived by considering the energetics taking place inside the stellar volume through the differential law
\begin{equation}
dL=\epsilon dm,
\end{equation}
where $\epsilon$ is the nuclear energy generation rate in units of $\rm erg\; g^{-1} \; s^{-1}$. In general, $\epsilon$ can display a rather complicate expression but its main contribution can be parametrized for a low mass solar-type star with active $pp$ chain under the form \citep{maciel}
\begin{equation}
\epsilon(r)= 2. 46 \times 10^6 \,\mbox{erg g}^{-1} \; \mbox{s}^{-1}\,  \rho(r) X^2 \left( \frac{T(r)}{10^6 \; \rm K}\right) ^{-\frac{2}{3}} e^{-33.81 \left( \frac{T(r)}{10^6 \; \rm K}\right) ^{-\frac{1}{3}}},
\end{equation}
where $X$ is the proton fraction and $T$ is the temperature of the star. Thus, the luminosity (in erg/s) can be derived as

\begin{equation}
L=\int_0^R 4\pi r^2 \rho(r) \epsilon(r) dr.
\end{equation}
 For most stars (with the exception of very low mass
stars and stellar remnants) the ions and electrons can be
treated as an ideal gas and quantum effects do not affect critically their behaviour. In our treatment and, in order to keep our modellization simplified, we will consider that the radiation pressure is much smaller than that of the gas of ions and electrons in the stellar plasma, i.e. $P_r\ll P_{\rm gas}$. Note that, for the particular example of  the sun core $P_r\sim 10^{-4}P_{\rm gas}$. As mentioned, a more general treatment \citep{MS1}  would involve considering a mixture of both pressure components. 

Therefore, in our scenario the stellar conditions are  dominated by the gas pressure $P\approx P_{\rm gas}=\frac{\rho k_B T}{\bar{\mu} m_H} $, with $\bar \mu$ the mean  molecular weight, $m_H$ the hydrogen mass and $k_B$ the Boltzmann constant. Then, the temperature can be written for the $n=3$ polytropic star as
\begin{equation}
T=\frac{\bar \mu m_H K \rho^{\frac{1}{3}}}{k_B}.
\label{temp}
\end{equation}
 
 \bigskip

\section{Results}

In this section we analyze the results obtained for several magnitudes of interest we have calculated in the context of the MG modellizations under study. In order to solve Eqs. \ref{ED1P} and \ref{ED2P}, we take $K=3.8 \; \rm {cm^3}{g^{-\frac{1}{3}}s^{-2}}$ so that we can recover typical values of mass and radius of solar-type stars, $R_0, M_0$ used as normalization. Although we do not attempt to accurately model stellar values of relevant magnitudes it is worth mentioning at this point that in our polytropic description, most of the star  i.e. in the region $r<0.7R_\odot$, is well approximated by a polytrope of index $n=3$ while for the outer region, with convective behaviour, it would be better described by a polytrope of index $n=1.5$. The latter corresponding, however, only to a 0.6$\%$ of the mass. 
\begin{figure}[ht]
\begin{center}
\includegraphics [angle=0,scale=0.75] {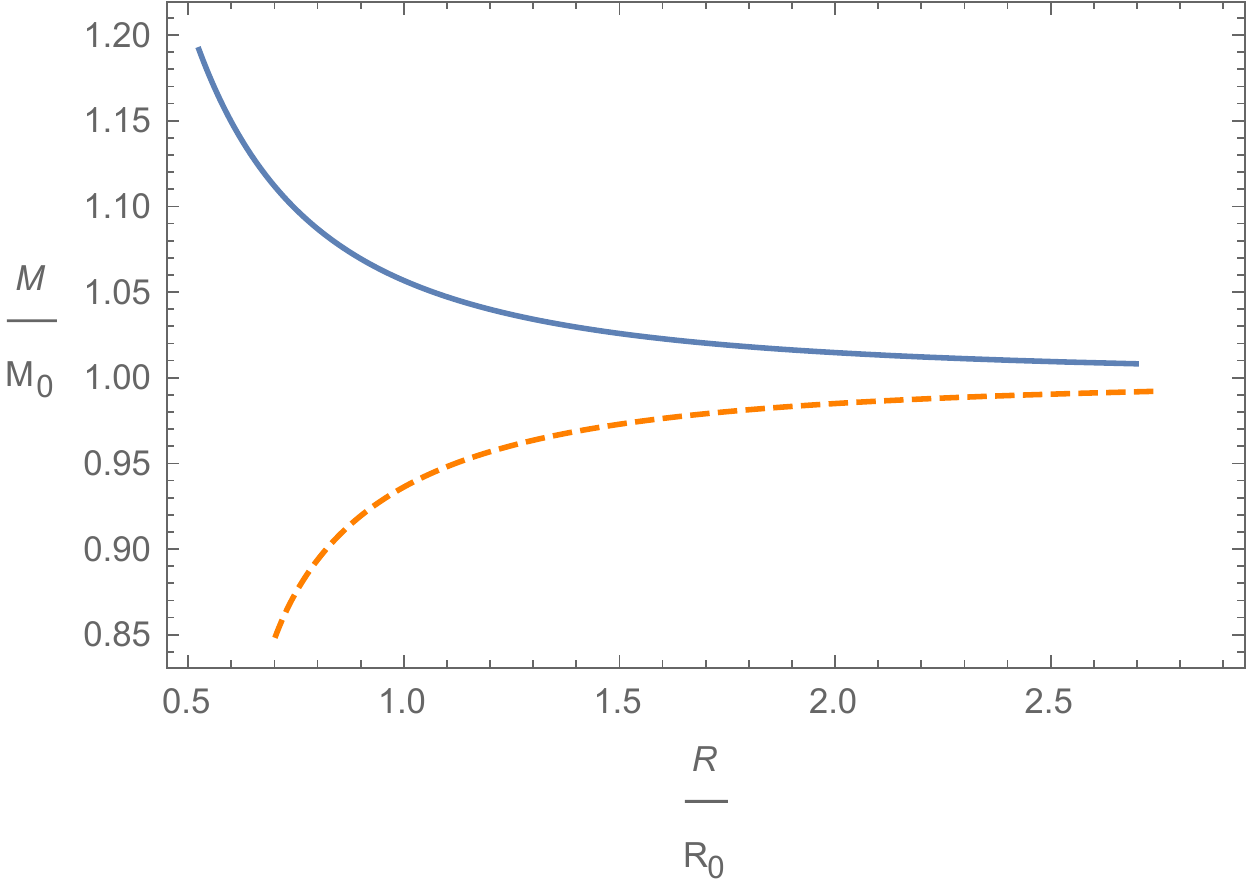}
\caption{Mass-Radius relationship for $|\xi|=4\times 10^{18} \; \rm cm^2$. Solid lines correspond to the case in which $\xi>0$ and dashed lines $\xi<0$. We use $\rho_c\in [0.05,5]\rho_0$. See text for details.}
\label{fig2}
\end{center}
\end{figure}
\begin{figure}[t]
\begin{center}
\includegraphics [angle=0,scale=0.75] {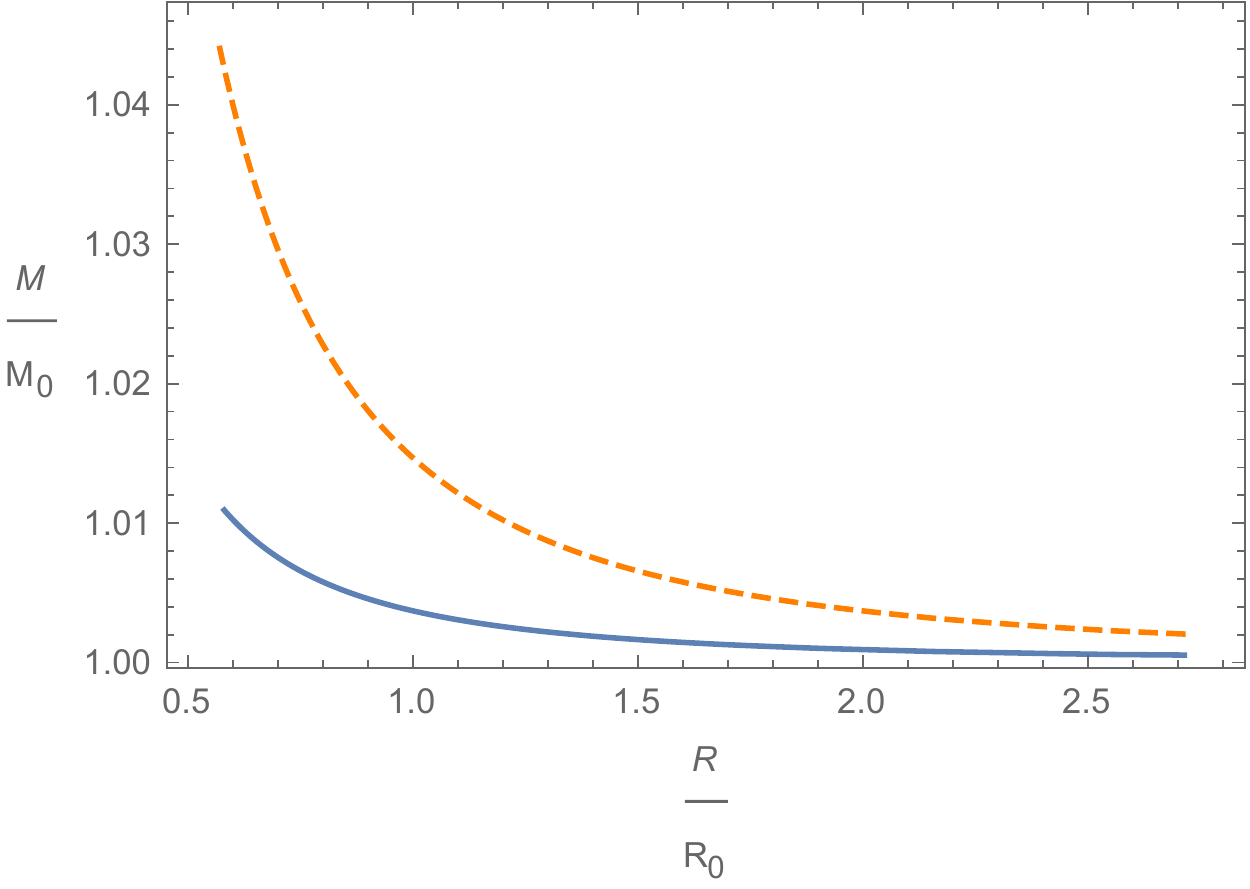}
\caption{Mass-Radius relationship for values $\xi=2.5\times 10^{17}\; \rm cm^2$ (solid lines) and $\xi=10^{18}\; \rm cm^2$ (dashed lines). We use $\rho_c\in [0.05,5]\rho_0$. See text for details.}
\label{fig6}
\end{center}
\end{figure}

In Fig.(\ref{fig1}) we consider the case of a low-density solar-type model with central density $\rho_c=\rho_0$ ( $\rho_0=80$ $\rm g/cm^3$). We 
show the solution potentials i.e. the potential before introducing the perturbation, $\psi_0$, and the perturbation, $\psi_1$, fulfilling $|\psi_1|\ll|\psi_0|$, as well as their sum $\psi=\psi_0+\psi_1$, as functions of the radial coordinate $r$ for $\gamma=1$ and $\xi=-2.67 \times 10^{18}$ cm$^2$. More generally, we obtain that only for values of $|\xi| \lesssim 1.225 \times 10^{19} \; \rm cm^2$ a perturbative correction $|\psi_1|\leq 0.1|\psi_0|$ indeed justifies our framework.

As we have mentioned before, the radius of the star in this approach can be obtained as the first zero of the full potential solution, $\psi(R)=0$. Once we know this value, the mass of the star is obtained as $M=M(R)=\int_0^R 4\pi r^2 \rho dr$. 

In Fig.(\ref{fig2}) we plot the $M-R$ relationship for $|\xi|=4\times 10^{18} \; \rm cm^2$. We normalize to values obtained in the solar-type case $R_0, M_0$. Solid lines correspond to the case in which $\xi>0$ and dashed lines $\xi<0$. We take $\rho_c\in [0.05,5] \rho_0$ to generate our data points. We can see that in all cases corresponding to MG the relation gets distorted from the $n=3$ GR case in which $M$ is constant when varying $\rho_c$. Objects along the M-R curve  with a positive slope $dM/dR>0$ denote metastable configurations so that in our analysis they are discarded.

In Fig.(\ref{fig6}) the $M-R$ diagram is shown. It has been obtained by varying the central density value in the interval $\rho_c\in [0.05, 5] \rho_0$. We use $\xi=2.5 \times 10^{17}\; \rm cm^2$ (solid lines) and $\xi= 10^{18}\; \rm cm^2$ (dashed lines). We can observe how the shape of the $M-R$ diagram slightly changes (compared to the flat $M/M_0=1$ result from GR) for different values of $\xi$, being more significant for objects with lower radii. The star will achieve more massive configuration as $\xi$ increases.

Stellar mass and radius can be described with an approximate fit as functions of $(\xi, \rho_c)$. In this way, for the stellar radius $R(\xi, \rho_c)$ is given by 
%
\begin{equation}
\frac{R}{R_0}=\left(1+\frac{\xi}{10^{19}\; \rm cm^2}\right)^{-0.116}\left( \frac{\rho_c}{\rho_{0}}\right)^{-0.333 - 0.075 \frac{\xi}{10^{19}
\; \rm cm^2}}.
\end{equation}
For masses $M(\xi, \rho_c)$, instead, the corresponding fit is 
\begin{eqnarray} 
 \frac{M}{M_0}=\left(1+\frac{\xi}{10^{19}\; \rm cm^2}\right)^{0.15} \left( \frac{\rho_c}{\rho_{0}}\right)^{0.1\frac{\xi}{10^{19}\; \rm cm^2}}.
 \end{eqnarray}
Using this fit GR solutions $M/M_0=1$ and $R/R_0=1$ are obtained when $\xi=0$ and ${\rho_c}=\rho_{0}$.
\begin{figure}[t]
\begin{center}
\includegraphics [angle=0,scale=0.7] {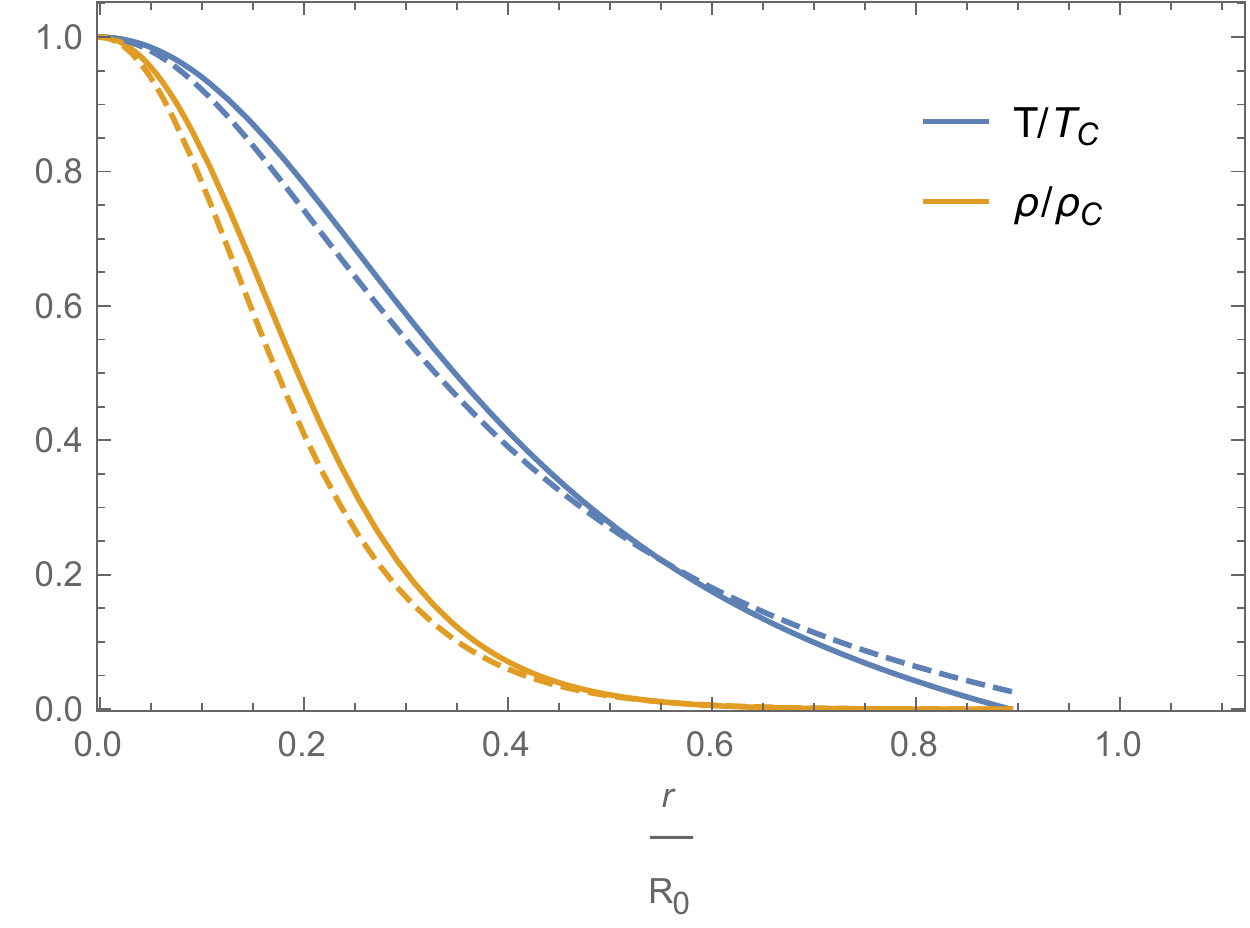}
\caption{Density and temperature as a function of $r/R_0$ for a solar-type star. We take $\xi=9\times 10^{18}\; \rm cm^2$ (solid lines) and $\xi=0$ (dashed lines) to compare with the GR case
}
\label{fig3}
\end{center}
\end{figure}

In Fig.(\ref{fig3}) we  plot stellar density and temperature (normalized to the central values) as a function of $r/R_0$, fixing $\rho_c=\rho_0$ for $\xi=9 \times 10^{18}\; \rm cm^2$ (solid lines) and $\xi=0$ (dashed lines), the latter corresponding to the GR case. Central temperature is obtained using Eq. \ref{temp} with $\bar \mu=0.61$, corresponding to Hydrogen, Helium fractions of solar-type stars $X\sim 0.7$, $Y\sim 0.3$ yielding a central value $T_c\sim 1.21 \times 10^7 \; \rm K$. As can be seen, the value of the density (temperature) profile for the MG case of $\xi=9 \times 10^{18}\; \rm cm^2$ is only slightly changed when compared to the GR case. The temperature radial profile, $T(r)$, also governs the change in the luminosity of (low mass) solar-type stars. 
\begin{figure}[t]
\begin{center}
\includegraphics [angle=0,scale=0.86] {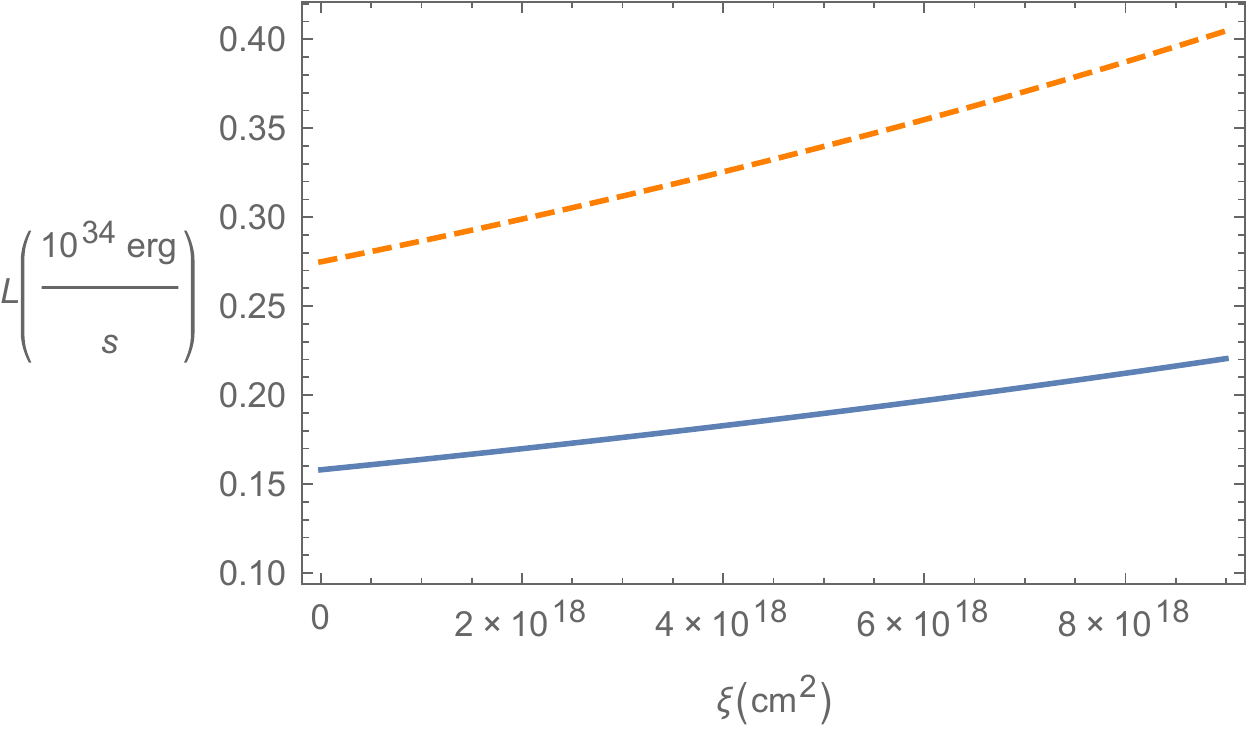}
\caption{Stellar luminosity for a solar-type star as a function of $\xi$ for two different values of  $\rho_c=\rho_0$(solid lines) and  $\rho_c=100$ $\rm g/cm^3$ (dashed lines). }
\label{fig4}
\end{center}
\end{figure}

In Fig.(\ref{fig4}) we  plot stellar luminosity  as a function of $\xi$. We consider $\rho_c=\rho_0$ (solid line) and  $\rho_c=100$ $\rm g/cm^3$ (dashed line).  From considerations relative to solar emission uncertainties, a $\sim \pm 1\%$ flux accuracy in the SSM could  accommodate variations of GR taking place below length-scales $\sqrt{\xi}\sim 10^9$ cm. Although variations expected for some objects may be indeed  larger it may result extremely difficult to disentangle the presence of such a component from a ordinary variation due to complex dynamics of stellar (solar) interior.

\begin{table}
\caption{\label{tab:table3} Slope values $\alpha(\xi)$ in the correlation $L$ versus $T_{\rm eff}$ for solar-type stars. See text for details.}
\begin{ruledtabular}
\begin{tabular}{lcr}
$\xi (\rm cm^2)$&$\alpha$\\
\hline
$-4\times 10^{18}$ & $3.234$ \\
$0$ & $3.1904$ \\
$4\times 10^{18}$ & $3.1702$\\
\end{tabular}
\end{ruledtabular}
\end{table}
\begin{figure}[t]
\begin{center}
\includegraphics [angle=0,scale=0.65] {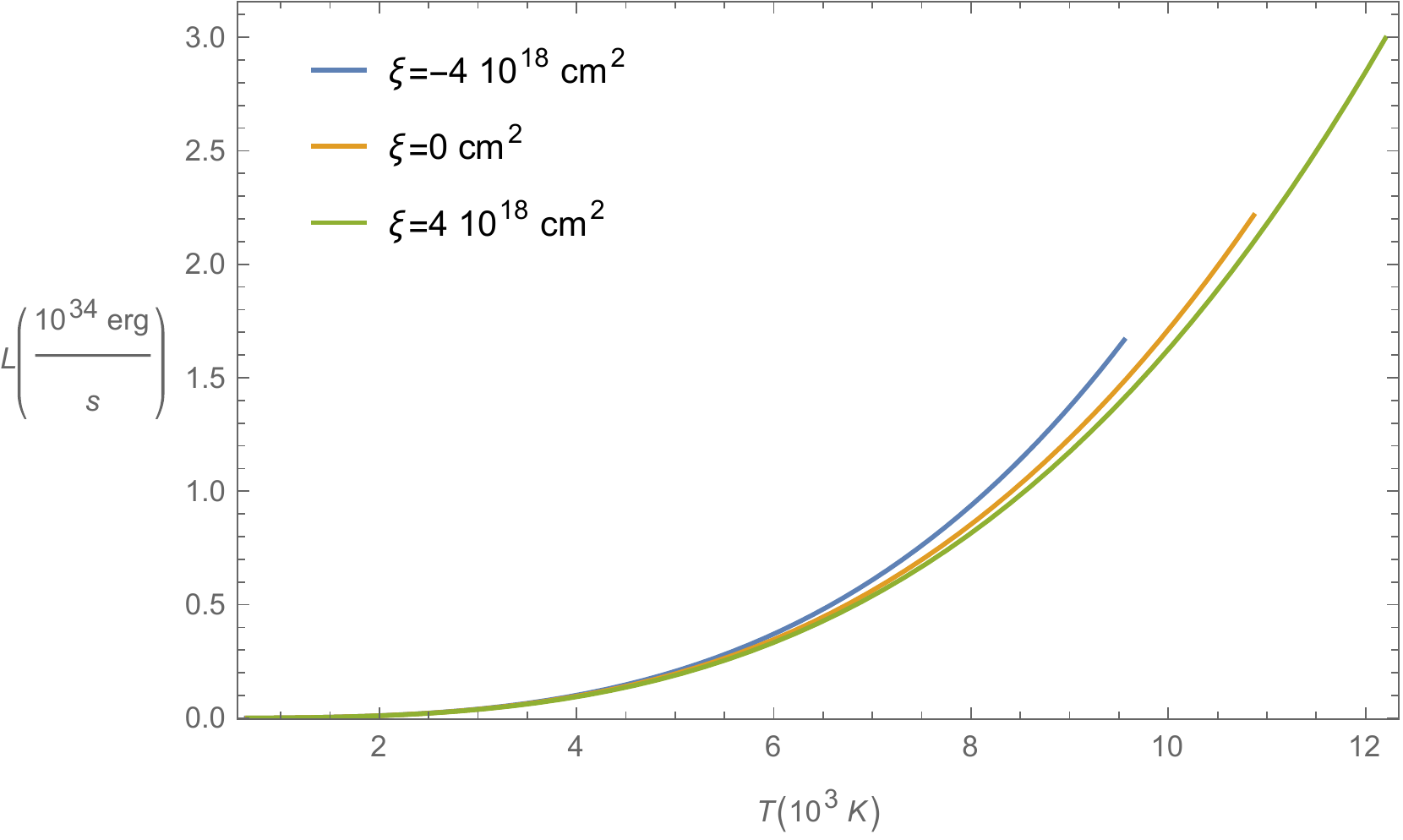}
\caption{Stellar luminosity for a solar-type star as a function of effective temperature $T_{\rm eff}$ for $\xi=-4\times 10^{18},\; 0$ (GR case) and $4\times  10^{18}$ cm$^2$. We take  $\rho_c\in [0.1, 3] \rho_0$.}
\label{fig5}
\end{center}
\end{figure}
In order to size the impact of the models under study in the MG scenario on the luminosity and  temperature a prescribed correspondence with a black-body spectrum $L=4 \pi R^2T^4_{\rm eff}$ has been used in Fig.(\ref{fig5}). We plot stellar luminosity as a function of effective temperature, $T_{\rm eff}$. 
We consider values $\xi=-4\times 10^{18}, 0$ and $4\times 10^{18}$ cm$^2$ (blue, orange and green lines, respectively). Values $\xi<0$ would correspond to metastable stellar configurations but they are shown for the sake of completeness.  We set values for the central density in the interval $\rho_c\in [0.1, 3] \rho_0$.  We can see there is a correlation of luminosity-temperature, as expected, and thus a non-trivial dependence on the effective parameter of our MG modellization, $\xi$, which results clear for $T_{\rm eff}>6000$ K. Spectral types O, B, A, F for main sequence stars can display such high temperatures. Although we can not expect to recover the meaningful (GR) Hertzsprung-Russell diagram with our simplified approach, the analogous logarithmic correlation we find under the form $\rm Log\,L=\alpha \rm Log\, [T_{\rm eff }(R)]+C$ yields a weak  variation in the slope, $\alpha(\xi)$, for this case at the $\sim 0.6\%$. See some values listed for reference in table \ref{tab:table3}. One important difference obtained in our calculation with respect to the usual approach where $(T_{\rm eff},R)$ are non-correlated variables is that for the GR case we do not recover the familiar value $\alpha=4$, since in our treatment $T_{\rm eff}$ is obtained from the R-dependent value of the luminosity. In other words, the used MG model determines the variation in the radius $R$ and thus, the luminosity. Therefore, values shown in table \ref{tab:table3} are by no means a predictive output of the model, but rather an indication of the weak dependence of the $\xi$ parameter.

\begin{figure}[t]
\begin{center}
\includegraphics [angle=0,scale=0.85] {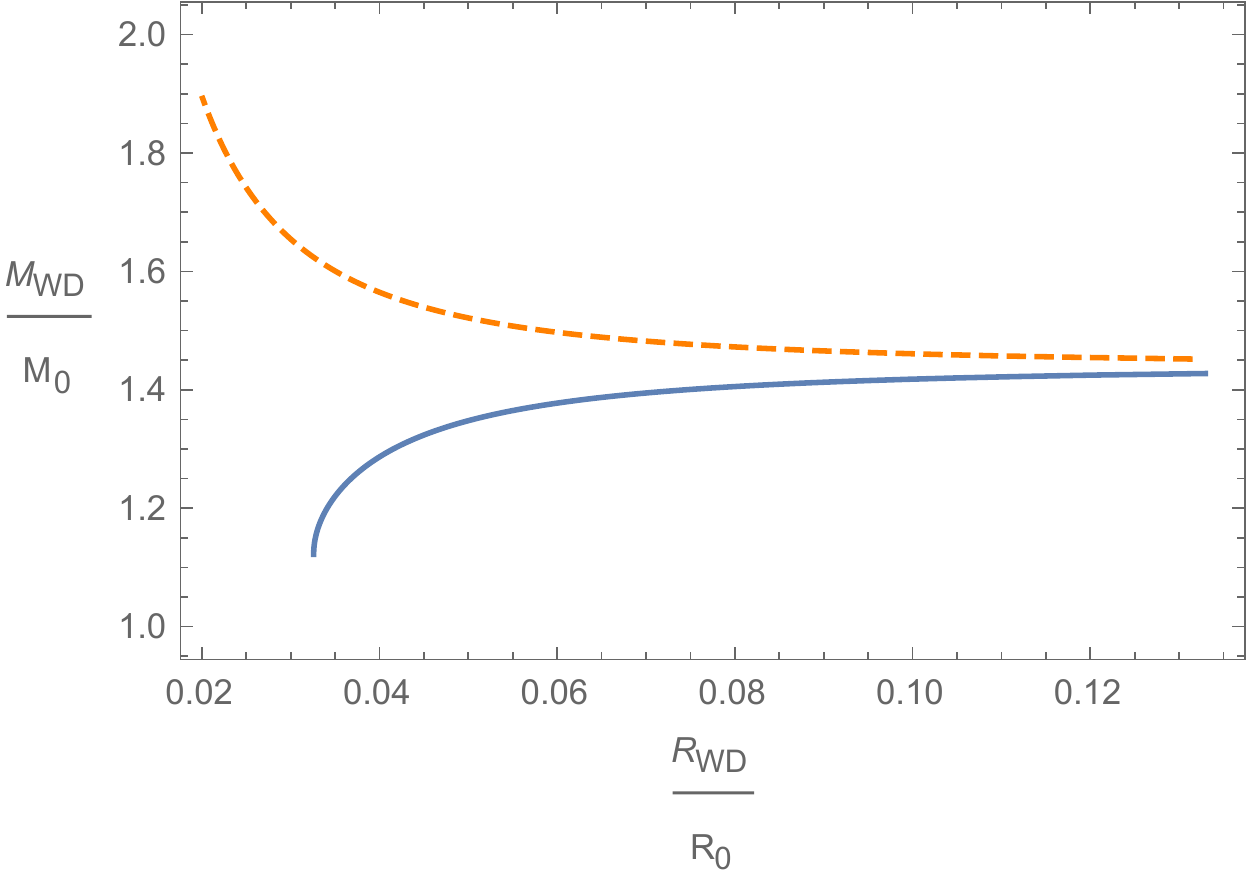}
\caption{ WD Mass-Radius relationship for $|\xi|=10^{16}\; \rm  cm^2$. Solid lines correspond to the case in which $\xi<0$ and for dashed lines $\xi>0$. We use $\rho_c\in [0.01,9]\rho_{c,I}$ where $\rho_{c,I}=10^6\,\rm g/cm^3$ .}
\label{fig90}
\end{center}
\end{figure}

In order to explore the variations introduced by the subset of MG model we study in this work let us consider a higher density stellar object such as a white dwarf (WD). To model this kind of objects we use a relativistic Fermi gas simplified EoS dominated by an electron component using a polytrope with n=3 and $K=4.8\times 10^{14} \; \rm cm^3 g^{-\frac{1}{3}}s^{-2}$. In this way we recover solutions to the structure equations Eqs. \ref{ED1P} and \ref{ED2P} yielding masses up to the maximum theoretical limit $M=1.44M_\odot$ (Chandrasekhar mass). Note that when using GR with a $n=3$ polytropic EoS, $M$ values (but not radii) are constant when varying $\rho_c$.

In Fig.(\ref{fig90}) we plot the WD mass-radius relationship for $|\xi|=10^{16}\; \rm  cm^2$. Solid (dashed) lines correspond to the case in which $\xi<0$ ($\xi>0$). We use  $\rho_c\in [0.01,9]\rho_{c,I}$ where $\rho_{c,I}=10^6\,\rm g/cm^3$ is a typical WD central density. As before, we keep for reference $M_0$ and $R_0$ for normalization in the axis labels using those given by solar-type stars.
If we further restrict to models in which $\xi \geq 0$, which are those which provide stable stellar configurations $M(R)$, it is clear that, given a $\xi$ value, the maximum mass ($M_{\rm Ch}$) will be obtained for the highest possible value of $\rho_c$ provided the perturbative solution still holds. Since the heaviest white dwarf reported in the literature has a mass $M/M_\odot \simeq 1.37$ \citep{chand}, and the GR $M_{\rm Ch}$ value is even higher than this, $\sim 1.44 M_\odot$, all values obtained in this perturbative framework are allowed. Thus in the high density setting presented the $\xi$ parameter remains unconstrained.

\section{Conclusions}

Using a model-independent approach, we have studied the effects of a subset of Modified Gravity theories involving one extra scalar degree of freedom on stellar structure. We use a perturbative technique capable of solving in a linearized regime. We have selected our main scenario concerning the application to low-density stellar objects. First, in a perturbative approach, we have obtained mass and radius solutions for low mass solar-type objects using a polytropic $n=3$ EoS. We have analyzed how internal temperature profiles and stellar luminosities are affected with respect to the reference case of GR. We provide a fit for masses and radii provided a central density value, i.e. $M(\xi,\rho_c)$,  $R(\xi,\rho_c)$. We obtain  that a change in curvature of $M(R)$ results when $\xi$ changes from positive to negative values, corresponding the latter to metastable stellar configurations. 
We also find that Modified Gravity can affect stellar luminosity from its $\xi$ dependence.  Globally, the effective temperature from a $L\propto T^4_{\rm eff}$ law results in a linear Log $L$-Log $T_{\rm eff}$ behaviour with a weak dependence in $\xi$. We anticipate  that this could result in objects appearing brighter (as L is increased with any stable configuration $\xi>0$) but seems hard to measure experimentally as internal stellar dynamics is complex.
Furthermore, it seems challenging to disentangle this effect from other standard effects such as proper fluctuations of the star. Even the Sun is a weakly variable star, with its major source of fluctuation coming from the eleven-year solar cycle and revealing a smaller periodic variation of about $\sim \pm 0.1\%$. It is nevertheless worth to point that both effects, involving standard and new physics, could indeed be present and need to be further studied.
When a particular case of a high density star is considered (white dwarfs) we obtain no constraint appears on the $\xi$ parameter from calculated values of the Chandrasekhar mass since it is allways larger than the GR value in the perturbative approach we use.  Our treatment for both low and high-density stellar case examples can help to understand the behaviour of Modified Gravity on small (not cosmological) scales.

\section*{acknowledgments}

We thank M. Fairbairn for useful comments. This work has been supported by Junta de Castilla y Le\'on SA083P17,  Spanish  MINECO grants FIS2015-65140-P,  FIS2016-78859-P(AEI/FEDER, UE)  and by PHAROS Cost action. We also thank the support of the Spanish Red Consolider MultiDark FPA2017-90566-REDC. M. Cerme\~no is supported by a fellowship from the University of Salamanca.

\end{document}